\pdfoutput=1
\newcommand*{\ATLASLATEXPATH}{./}
\documentclass[cernpreprint, USenglish,texlive=2016,txfonts]{\ATLASLATEXPATH atlasdoc}

\usepackage{\ATLASLATEXPATH atlaspackage }
\usepackage{\ATLASLATEXPATH atlasbiblatex}

\usepackage{\ATLASLATEXPATH atlascontribute}

\usepackage{\ATLASLATEXPATH atlasphysics}

\usepackage{multicol}
\usepackage{multirow}
\usepackage{url}
\usepackage{wrapfig}
\usepackage{indentfirst}
\usepackage{siunitx} 
\newboolean{Paper}
\setboolean{Paper}{true}

\makeatletter
\newcommand*{\numtablerange}[1][]{%
  \kernel@ifnextchar[{\numtable@range{#1}}{\numtable@range{#1}[#1]}}

\def\numtable@range#1[#2]#3#4{%
  \numtablerange@boxme{\expandafter\tablenum\expandafter[\numtablerange@lopt, #1]{#3}} $\pm$
  \numtablerange@boxme{\expandafter\tablenum\expandafter[\numtablerange@ropt, #2]{#4}}
}

\newcolumntype{T}[3]{>{\def\numtablerange@lopt{#1}\def\numtablerange@ropt{#2}}#3}

\newcommand*{\numtablerange@boxme}[1]{{\sbox0{#1}\usebox0}}
\makeatletter

\newcommand{\ilumi} {$ 19.5\, \mathrm{fb}^{-1} $}
\newcommand{\ilumiSeven} {$ 4.9\, \mathrm{fb}^{-1} $}

\newcommand{\CoMEnergyEight}{$ 8~\TeV$}
\newcommand{\CoMEnergySeven}{$ 7~\TeV$}

\newcommand{\rhoATLAS}{\ensuremath{\rho_X}}

\newcommand{\NXten} {$N(X) < 382$  }
\newcommand{\NXfif} {$N(X) < 356$  }

\newcommand{\rhoten} {$\rhoATLAS <  0.015$  }
\newcommand{\rhofif} {$\rhoATLAS <  0.016$  }

\newcommand{\cutChiBs}   {$\chi^{2} /\mathrm {d.o.f.} < 3$}

\newcommand{\cutBspTten} {$ \psubt(\Bs)>10~\GeV $}
\newcommand{\cutBspTfif} {$ \psubt(\Bs)>15~\GeV $}

\newcommand{\Bspi}{\ensuremath{ \Bs~\pi^\pm }} 
 
\newcommand{\Xt}{\ensuremath{ X \ra  \Bs~\pi^\pm }} 

\newcommand{\Bst}{\ensuremath{ \Bs \ra \JpsiPhi }}

\newcommand{\Ds}{\ensuremath{D_{s}} \space}

\newcommand{\Jpsipa}{\Jpsi \ }

\newcommand{\psubt}{\pt \space}

\newcommand{\Jpsitomm}{\ensuremath{ \Jpsi \ra \mumu }\space} 
\newcommand{\phiKK}{\ensuremath{ \phi \ra \kplus \kminus}\space}

\newcommand{\BstoDsmu}{\ensuremath{ \Bs \ra \mu^{\mp} \nu \Ds^{\pm} }}

\newcommand{\Dst}{\ensuremath{ \Ds^{\pm} \ra \phi \pi^{\pm}  }}

\newcommand{\fb}{\ensuremath{fb^{-1}}}
\newcommand{\nb}{\ensuremath{nb^{-1}}}

\newcommand{\JpsiPhi}{\ensuremath{ J/\psi\phi }}
\newcommand{\kplus}{\ensuremath{K^+}} 
\newcommand{\kminus}{\ensuremath{K^-  }}

 \usepackage{authblk}
 
 \AtlasTitle{Search for a Structure in the \Bspi \ Invariant Mass Spectrum with the ATLAS Experiment}

 \author{The ATLAS Collaboration}
\PreprintIdNumber{CERN-EP-2017-333}

\AtlasJournalRef{\PRL 120 (2018) 202007}
\AtlasDOI{10.1103/PhysRevLett.120.202007}

\AtlasAbstract{%
A search for the  narrow structure, $X(5568)$, reported by the D0 Collaboration in the decay sequence  \Xt, \ \Bst, is presented.
The analysis is based on a data sample recorded with the ATLAS detector at the LHC corresponding to    \ilumiSeven \ of  $pp$ collisions at \CoMEnergySeven \ and    \ilumi \  at \CoMEnergyEight.  No significant signal was found. 
Upper limits on  the number of signal events, with properties  corresponding to those reported by  D0,  and  on the $X$ production rate  relative to \Bs mesons,  \rhoATLAS,  were determined at 95\% confidence level. The results are  \NXten \ and \rhoten \  for $B^0_s$ mesons with transverse momenta above $10~\GeV$, \ and \NXfif \ and \rhofif \  for transverse momenta above $15~\GeV$. Limits are also set for potential \Bspi \ resonances in the mass range  $5550~\MeV$  to  $5700~\MeV$. 
 }

\hypersetup{pdftitle={ATLAS document},pdfauthor={The ATLAS Collaboration}}

\begin{document}

\maketitle

\clearpage





The D0 Collaboration reported evidence of a narrow structure, $X(5568)$, in the decay $\Xt$ with $\Bst$
in proton--antiproton collisions at a center-of-mass energy of
$\sqrt{s}=1.96~\TeV$
at the Tevatron collider~\cite{D0bspi}.
The structure was interpreted as
a tetraquark with four different quark
flavors: $b$, $s$, $u$ and $d$.
The mass and natural width of this state were fitted to be
$m = 5567.8\pm 2.9 (\textrm{stat})^{+0.9}_{-1.9} (\textrm{syst})~\MeV$ and
$\Gamma=21.9\pm6.4 (\textrm{stat})^{+5.0}_{-2.5} (\textrm{syst})~\MeV$,
respectively, and the signal significance is 5.1~$\sigma$.
The ratio $\rho_X$ of the yield of $X(5568)$ to the yield
of the \Bs meson
for a transverse momentum range $10<\pt(\Bs)<30~\GeV$
was measured to be
$0.086\pm 0.019 (\textrm{stat})\pm 0.014 (\textrm{syst})$.
The result initiated a  discussion of the nature of the new
state and prospects for observation of other
tetraquark hadrons~\cite{WeiWang,WeiChen,YRLiu,ZGWang,FlStancu}.
Recently, the D0 Collaboration reported further evidence for the resonance $X(5568)$ ~\cite{D0bsDs}  in the decay sequence  \Xt,  \ \BstoDsmu, \  \Dst,  which is consistent with their previous measurement ~\cite{D0bspi}.  However, searches for $X(5568)$ in decays to $\Bspi$, \Bst \ performed by the LHCb~\cite{LHCbbspi}
and CMS~\cite{CMSbspi} Collaborations in proton--proton ($pp$) collisions at the LHC and by the CDF Collaboration~\cite{CDFbspi} at the  Tevatron,  revealed no signal.
The  upper limits $\rho_X<0.024$ (LHCb, $\pt(B_s^0)>10~\GeV$), $\rho_X<0.011$ (CMS, $\pt(B_s^0)>10~\GeV$) and $\rho_X<0.010$ (CMS, $\pt(B_s^0)>15~\GeV$)
at $95\%$ confidence level (CL) were determined within the acceptances of the LHCb and CMS experiments. CDF  set an upper limit $\rho_X<0.067$ at $95\%$  CL 
 within a kinematic range similar to that of D0~\cite{D0bspi}.



\begin{figure}[t]
\begin{center}

\includegraphics[width=0.70\textwidth]{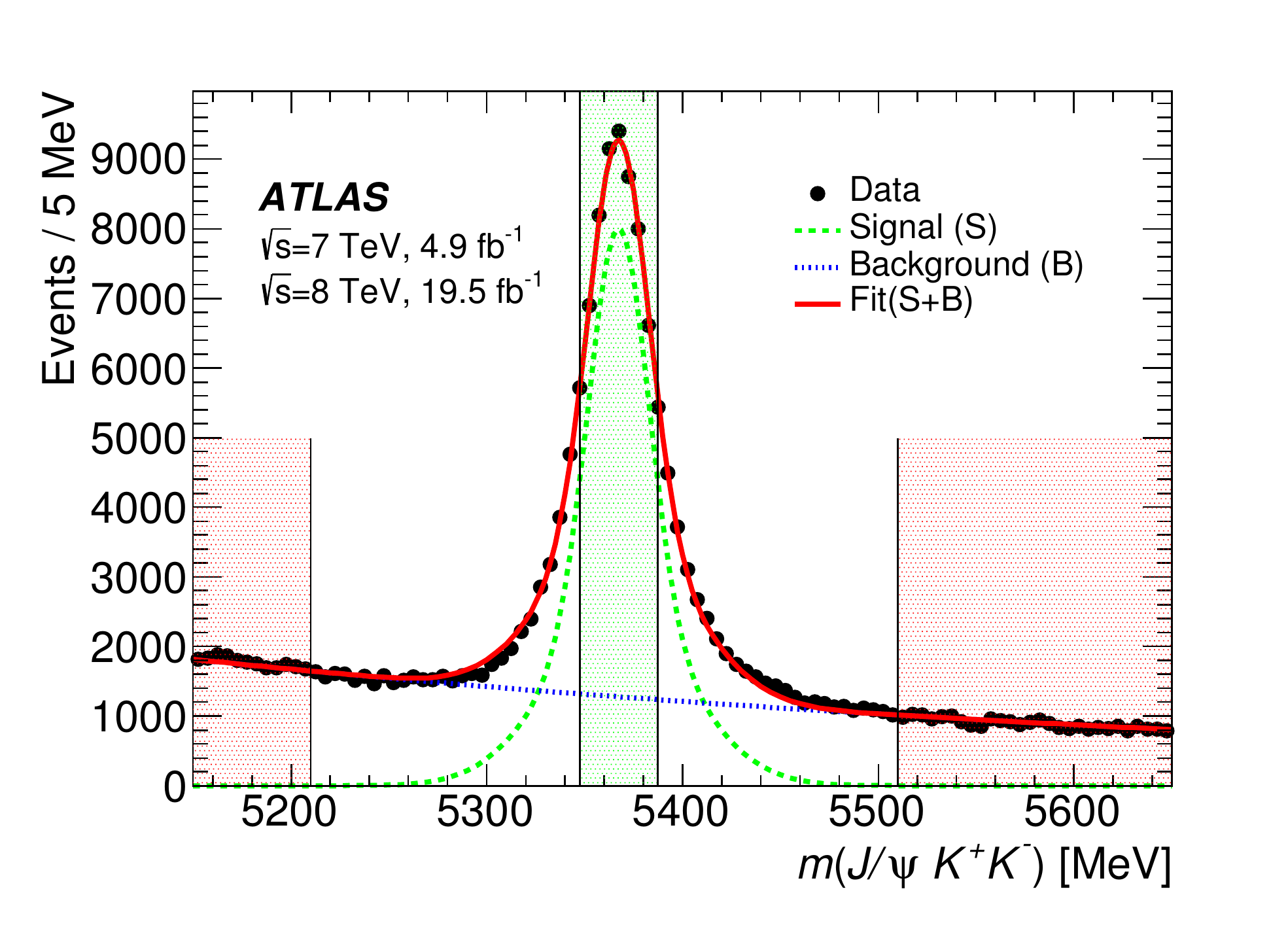}
                                                                     
\caption{The invariant mass distribution for   \Bst \ candidates satisfying the selection criteria.
Data are shown as points and results of fits to signal (dashed), background (dotted), and the total fit (solid) are shown
as lines. The two outer (red) shaded bands and the central (green) shaded band represent the  mass sidebands and the signal region of  \Bs meson candidates, respectively.}
\label{Fig:BsMass}
\end{center}
\end{figure}

In this Letter, a search for the $X(5568)$ state
by the ATLAS experiment at the LHC is presented ( \Bs  
refers to both the \Bs and ${\bar B}^0_s$ mesons).
The \Bs mesons are reconstructed in their decays to
$J/\psi(\mu^+\mu^-)\,\phi(K^+K^-)$.
The analysis is based on a combined sample of $pp$ collision data at $\sqrt{s}=7~\TeV$ and $8~\TeV$
corresponding to integrated luminosities of $4.9\,$fb$^{-1}$ and $19.5\,$fb$^{-1}$, respectively.
The ATLAS detector~\cite{ATLAS} covers nearly the entire solid angle around the collision point with layers of tracking detectors, calorimeters and muon chambers. The muon and tracking systems are of particular importance in the reconstruction of $B$ mesons. The inner tracking detector (ID) consists of a silicon pixel detector, a silicon microstrip detector and a transition radiation tracker. The muon spectrometer (MS) surrounds the calorimeters and consists of three large superconducting toroids with eight coils each, a system of  tracking chambers, and detectors for triggering.
To study the detector response, to estimate backgrounds and to model systematic effects, 12 million Monte Carlo (MC) simulated \Bst\ and  1 million \Bspi \ events were generated using  {\scshape Pythia} 8.183~\cite{Pythia, Sjostrand:2007gs} tuned with  ATLAS data~\cite{ATL-PHYS-PUB-2011-009}.
Multiple overlaid proton--proton collisions (pileup) were simulated with
{\scshape Pythia} soft QCD processes.
The detector response was simulated using the ATLAS simulation framework~\cite{Atlfast} based on GEANT4~\cite{Geant4}.  
The MC events were weighted to reproduce the same pileup and trigger conditions as in the data. 
 As in the D0 analysis~\cite{D0bspi},  the  \Bspi \ resonance was generated   using the Breit--Wigner (BW) parameterization appropriate for an $S$-wave two-body decay near threshold:

\begin{equation}
  F_\textrm{BW} \left( m(\Bspi), m_X, \Gamma_X \right) = \frac{ m(\Bspi) \cdot m_X \cdot \Gamma \left( m(\Bspi), \Gamma_X \right) } { \left( m^2_X - m^2(\Bspi) \right)^2 + m^2_X \cdot \Gamma^2 \left( m(\Bspi), \Gamma_X \right)},
  \label{Eq:BWSwave}
\end{equation}

 where $m(\Bspi)$ is the invariant mass of the \Bspi \ candidate and $m_X$ and $\Gamma_X$ are the mass and the natural width of the resonance. 
 The mass-dependent width is $\Gamma \left( m(\Bspi), \Gamma_X \right) = \Gamma_X \cdot (q_1/q_0)$,  where $q_1$ and $q_0$ are the magnitudes of the three-vector momenta of the  \Bs meson in the rest frame of the \Bspi \ system at the invariant masses equal to $m(\Bspi)$ and $m_X$, respectively.
The mass and the width were set to $m_X = 5567.8~\MeV$ and $\Gamma_X = 21.9~\MeV$, as reported in Ref.~\cite{D0bspi}.  
The events were selected  by the dimuon triggers~\cite{TRIG-2012-03} based on identification of a \Jpsitomm decay, with \pT
thresholds of either  $4~\GeV$ or $6~\GeV$, with both symmetric, $(4,4)~\GeV$ or  $(6,6)~\GeV$, and asymmetric, $(4,6)~\GeV$, combinations. 
 In  addition, each event must contain at least one reconstructed primary vertex (PV), formed from at least six ID tracks.
 The selection of  \Jpsi\ and \phiKK  candidates is identical to the one described in detail in Ref.~\cite{BPHY-2014-03}.
 Candidates for \Bst \ decays are selected by fitting the tracks for each combination of \Jpsitomm\ and \phiKK to a common vertex.
The fit is further constrained by fixing the invariant mass of the two muon tracks to the \Jpsipa mass~\cite{PDG16}. A quadruplet of tracks is accepted for further analysis if the vertex fit has a \cutChiBs. 
For each \Bs\ meson candidate the proper decay time $t$ is extracted  using the method described  in Ref.~\cite{BPHY-2014-03}.
Events with $t$~> 0.2~ps are selected to reduce the background from the events with a \Jpsi\ produced directly in the $pp$ collision.   If there is more than one accepted  \Bs\  candidate in the event, the candidate with the lowest $\chi^2 /$d.o.f. of the vertex fit is selected.  
For the selected events the average number of proton--proton interactions per bunch crossing is 21, necessitating a choice of the best candidate for the PV at which the \Bs\  meson is produced. 
The variable used is the three-dimensional impact parameter $d_0$, which is calculated as the distance between the line extrapolated from the reconstructed  \Bs\  meson vertex in the direction of the  \Bs\  momentum, and each PV candidate. The chosen PV is the one with the smallest $d_0$. Using MC simulation it was shown that the fraction of \Bs\ candidates that are assigned the wrong PV is less than 1\%~\cite{BPHY-2014-03} and that the corresponding effect on the results is negligible. 
Finally, a requirement that the \Bs\ transverse momentum is greater than $10~\GeV$  is applied. 
Figure \ref{Fig:BsMass} shows the reconstructed $J/\psi K^+K^-$ mass distribution and the result of an  extended unbinned maximum-likelihood fit  in the range  (5150 -- 5650)$~\MeV$, 
in which the signal is modeled by a sum of two Gaussian distributions and an exponential function is used to model the combinatorial background. The observed signal width is consistent with MC simulation. The fitted \Bs mass is $m_\mathrm{fit}(\Bs)$ = 5366.6~$\pm$~0.1~(stat)$~\MeV$, in agreement with the world average value 5366.89~$\pm$~0.19$~\MeV$~\cite{PDG16}.  For further investigation, only candidates with a reconstructed mass in the signal region 5346.6 -- 5386.6$~\MeV$ are included, which gives $N(B_s^{\,0})$ = 52750~$\pm$~280~(stat) candidates.

\begin{figure}[t]
  \centering
  \includegraphics[width=0.49\textwidth]{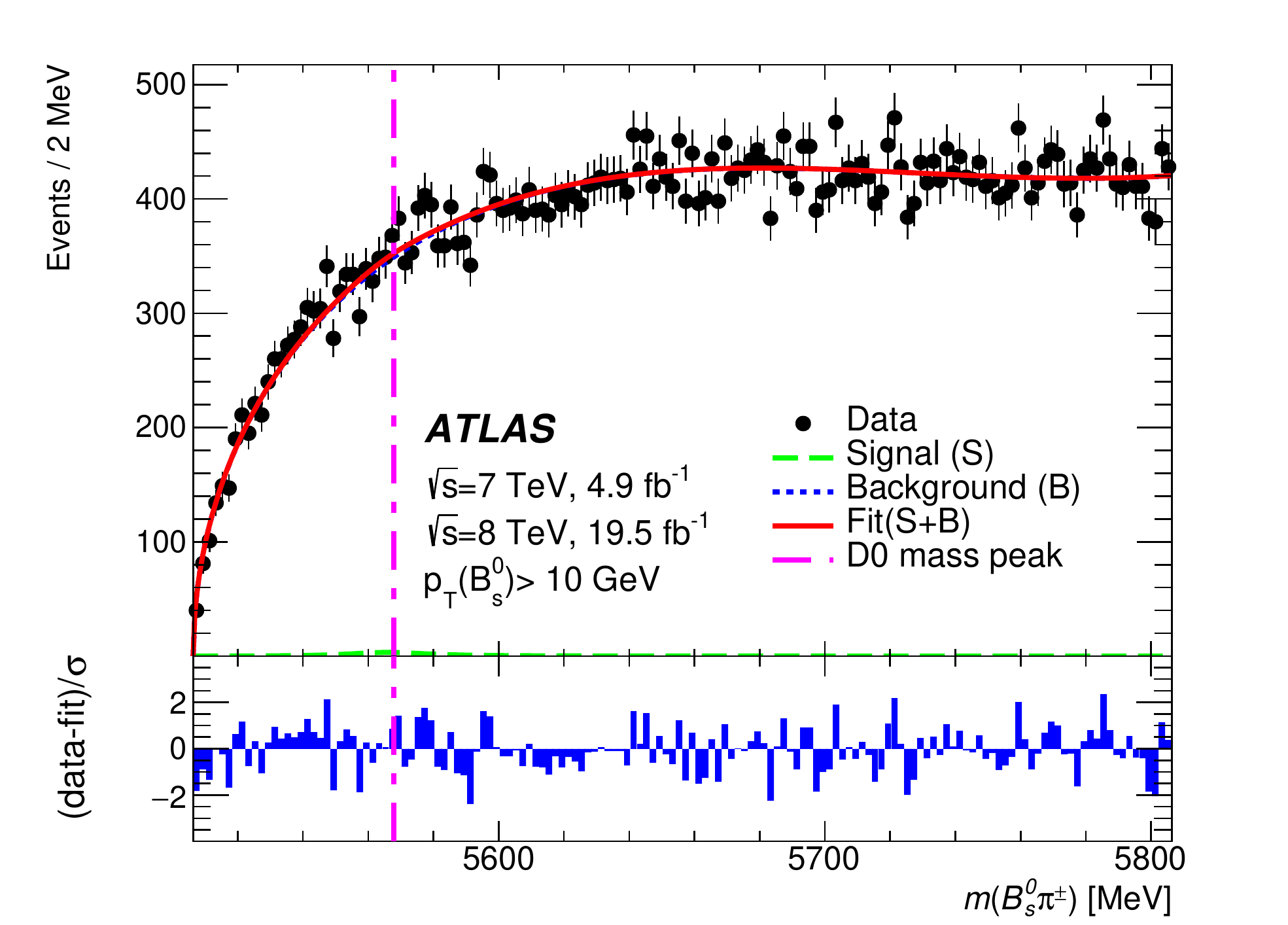}
  \includegraphics[width=0.49\textwidth]{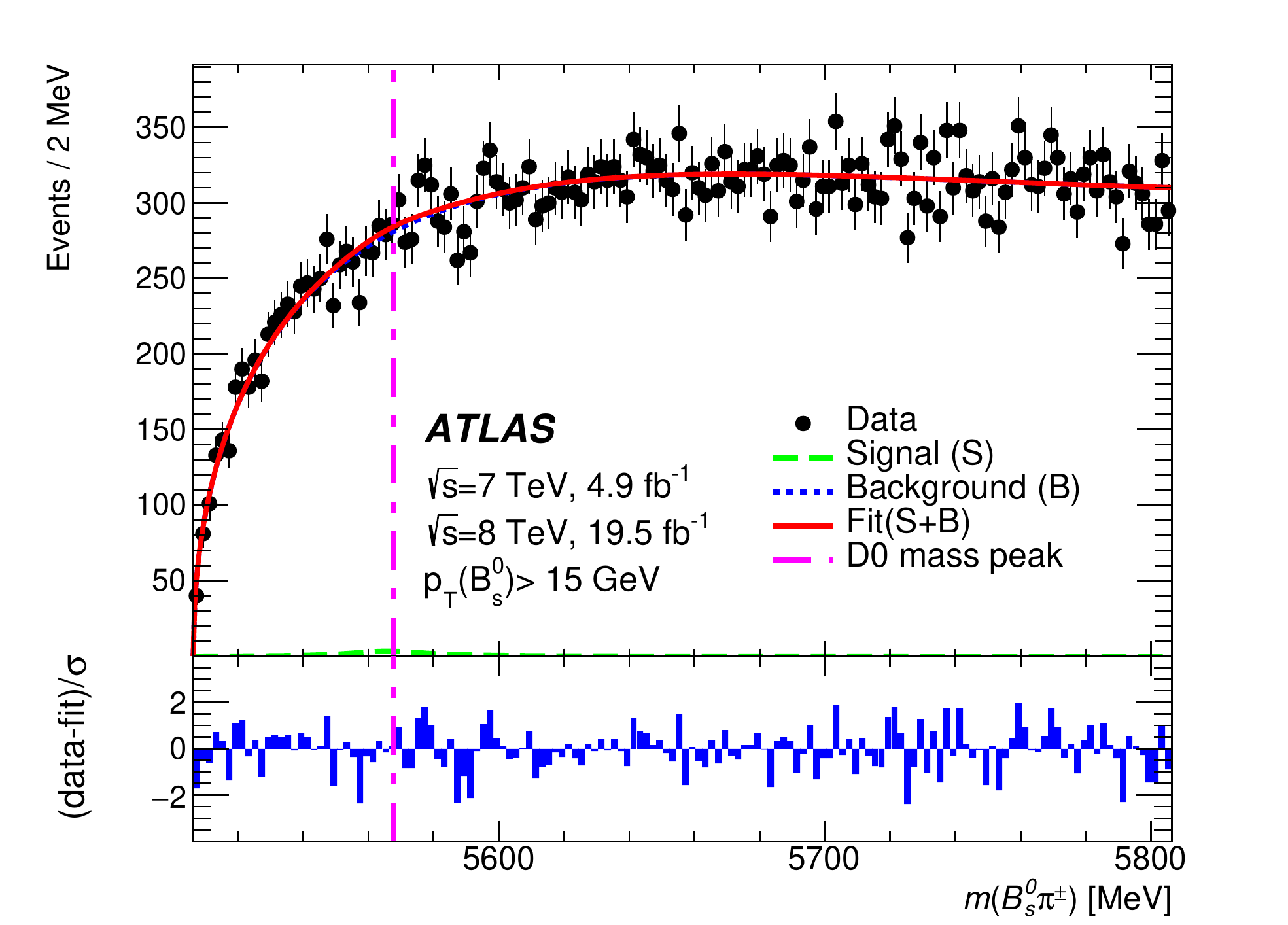}
  \caption{Results of the fit to the \Bspi \ mass distribution for candidates with \cutBspTten \ (left) and \cutBspTfif \ (right). 
   The bottom panels show the difference between each data point and the fit divided by the statistical uncertainty of that point.}
  \label{Fig:BsPiFit}
\end{figure}

The  \Bspi \ candidates are constructed by combining each of the tracks forming the selected PV with the selected \Bs candidate. Tracks that were already used to reconstruct the \Bs candidate and
tracks identified as leptons ($e$ or $\mu$) are excluded, as well as tracks with transverse momentum $\pT < 500~\MeV$.  This   \pT selection was chosen  to  maximize  the ratio of the $\Bspi$ signal to the background, based on MC simulation.  
Assigning the pion mass hypothesis to the tracks that pass these selection criteria, the  mass $m$(\Bspi)   is calculated as $m(J/\psi KK  \pi^\pm) - m( J/\psi KK ) + m_\mathrm{fit}(\Bs)$, where  $m_\mathrm{fit}(\Bs)$~ = ~ 5366.6$~\MeV$.
 On average there are 1.8~\Bspi \ candidates  in each selected event and all are retained for the analysis. A systematic study has shown that the effect on the results due to multiple candidates is negligible.
The mass distribution of $\Bspi$ candidates is fitted using an  extended unbinned maximum-likelihood method.
The probability density function (PDF) for the background component is defined as a threshold function:

\begin{equation}
  F_\mathrm{bck} \left( m(\Bspi) \right) = \left(\frac{m(\Bspi) - m_\mathrm{thr}}{n}\right)^a \cdot \exp{\left(\sum_{i=1}^4{p_i \cdot \left(\frac{m(\Bspi) - m_\mathrm{thr}}{n}\right)^i}\right)} ,
  \label{Eq:Fbck}
\end{equation}

where $m_\mathrm{thr} =   m_\mathrm{fit}(\Bs) + m_\mathrm{\pi}$ and $n$, $a$, $p_i$ are free parameters of the fit. The  background PDF was tested using events with no real \Bspi \ candidates  from two categories. The first background sample contains data events where \Bspi \ candidates are formed using ``fake'' \Bs mesons from the mass sidebands, shown in Figure~\ref{Fig:BsMass} by red shaded bands, 
defined as $5150 < m(J/\psi K^+K^-) < 5210~\MeV$ and $5510 < m(J/\psi K^+K^-) < 5650~\MeV$. The second background sample is modeled using MC events containing only  \Bs mesons not originating from  the  \Bspi \ signal, tuned  to reproduce the \Bs  transverse momentum distribution in data.
In these events the  \Bs meson  is   combined with each of the tracks originating from the selected PV. The first sample is normalized to the fitted number of \Bs background events in the  \Bs mass signal region 5346.6 -- 5386.6$~\MeV$, while the second sample is  normalized to the fitted number of \Bs signal events in the same region. The sum of these two distributions is consistent with the distribution of the data. The function in Eq.~(\ref{Eq:Fbck}) describes both background distributions as well as their sum within uncertainties.
The signal PDF $F_\mathrm{sig} \left(m(\Bspi)\right)$  is defined as a convolution of an $S$-wave Breit--Wigner PDF, defined in Eq.~(\ref{Eq:BWSwave}),  and the detector resolution 
 represented by a Gaussian function with a width that is calculated individually for each \Bspi \  candidate from the tracking and vertexing error matrices. 
  Using MC and data samples, it has been verified that the per candidate mass resolutions are the same for the  \Bspi \ signal and for the background events passing the selection criteria. 
 The average resolution  for the \Bspi \ signal, with the mass and width corresponding to those of the structure reported by the D0 Collaboration ($m_X = 5567.8~\MeV$ and $\Gamma_X = 21.9~\MeV$),  is  $3.2~\MeV$. The full probability function used is
   
\begin{equation}
  F\left(m(\Bspi)\right) = N(X) \cdot F_\mathrm{sig}\left(m(\Bspi)\right) + (N_\mathrm{can} - N(X)) \cdot F_\mathrm{bck}\left(m(\Bspi)\right),
   \label{Eq:FTotalFit}
\end{equation}

where $N(X)$ is the number of signal events and $N_\mathrm{can}$ is the number of all selected \Bspi \  candidates. The signal mass and width are fixed to the central values reported by the D0 Collaboration.
Following other experiments, fits are performed for  two subsets of \Bspi \ candidates, first with  \cutBspTten \ and second with \cutBspTfif.  The results of the fits are shown in Figure~\ref{Fig:BsPiFit} and summarized in Table~\ref{Tab:BsPiiFit}. No significant $X(5568)$ signal is observed. Additional selections such as cuts on the angle between the momenta of the  \Bs and  $\pi^\pm$ candidates were investigated and did not produce evidence of a signal. These were found to introduce peaking background so are not included in the analysis.
The yields $N(X)$ and $N(\Bs)$ obtained from the fits  are used to evaluate the $X$ production rate relative to $\Bs$, within the ATLAS acceptance, using the formula

\begin{equation}
\rhoATLAS \equiv \frac{\sigma(pp \rightarrow X + \mathrm{anything}) \times \mathcal{B}(X\rightarrow \Bs\pi^{\pm})}{\sigma(pp\rightarrow \Bs + \mathrm{anything})}
=\frac{N(X)}{N(\Bs)}\times\frac{1}{\epsilon^{\mathrm{rel}}(X)},
\label{Eq:Rho}
\end{equation}
where  $\sigma$  represents the production cross-section for each of the particles, within the ATLAS acceptance,
and
the relative efficiency $\epsilon^{\mathrm{rel}}(X) = \epsilon(X)/\epsilon(\Bs)$
is the selection efficiency for the state $X$, decaying to $\Bspi$,
relative to that for the $\Bs$ meson and accounts for the reconstruction and selection efficiency of the companion pion, including the soft pion acceptance.
\begin{table}[ht]
  \centering
\caption{Yields of \Bs and $X(5568)$ candidates obtained from the fits to the \Bs and $\Bs\pi^\pm$ candidate mass distributions, with statistical uncertainties. The values given for $N(\Bs)$ are those inside the \Bs signal window. The reported values for $X(5568)$ are obtained from  the fits with signal mass and width parameters fixed to those reported by the D0 Collaboration. The relative efficiencies $\epsilon^\mathrm{rel}(X)$ and their uncertainties are described  in  the text.\\
 } 
  \begin{tabular}{| c  c  c |}
  \hline 
  \multirow{2}{*}{$N(\Bs)/10^3$} & \cutBspTten & $\num{52.75 \pm0.28}$ \\
                                   & \cutBspTfif & $\num{43.46 \pm0.24}$ \\
  \hline
  \multirow{2}{*}{$N(X)$} & \cutBspTten & $ ~~~~60 \pm140$ \\
                          & \cutBspTfif & $~~-30 \pm150$ \\
                          
   \hline
  \multirow{2}{*}{$\epsilon^\mathrm{rel}(X)$} & \cutBspTten & $~~~\num{0.53 \pm0.09}$ \\
                                             & \cutBspTfif & $~~~\num{0.60 \pm0.10}$ \\

  \hline
  \end{tabular}
   \label{Tab:BsPiiFit}
\end{table}

The relative efficiency, $\epsilon^{\mathrm{rel}}(X)$, was determined using MC simulation of events containing $\Xt$ and $\Bs$ decays. In the ratio, the acceptance of the  \Bs decay cancels,  so the value to be determined  is the pion reconstruction efficiency for \Bspi \ events in which the \Bs \  meson satisfies acceptance, reconstruction and selection criteria. 
Based on MC events,   $\epsilon^{\mathrm{rel}}(X)$  is  determined as a function of $\psubt(\Bs)$ and of  $m$(\Bspi). 
Using an MC-based function, the acceptance is determined individually for each \Bspi \ candidate,
based on its measured values of $\psubt(\Bs)$ and $m$(\Bspi).
The acceptance ratio, $\epsilon^{\mathrm{rel}}(X)$, is calculated as an average over the events included in the $m$(\Bspi) interval within which the search for a resonance is performed. The width of this interval is defined by a BW function convolved with the mass resolution function, with the start and end points of the range chosen to include 99\% of the signal events. The uncertainty of $\epsilon^{\mathrm{rel}}(X)$ is calculated by varying the fitted parameters of the MC-based function used to describe the acceptance as a function of  $\psubt(\Bs)$ within their uncertainties. Small variations of this
function due to the pseudorapidity of the  \Bs \  were investigated and are included in the systematic uncertainties. The error also includes the uncertainty in the number of data events used in the average and the statistical uncertainty in the $\psubt(\Bs)$  distribution of these events.  The error in the pion reconstruction efficiency, arising from uncertainties in the amount of ID material, is found to have a negligible effect on  \rhoATLAS.

\begin{figure}[t]
  \centering
  \includegraphics[width=0.80\textwidth]{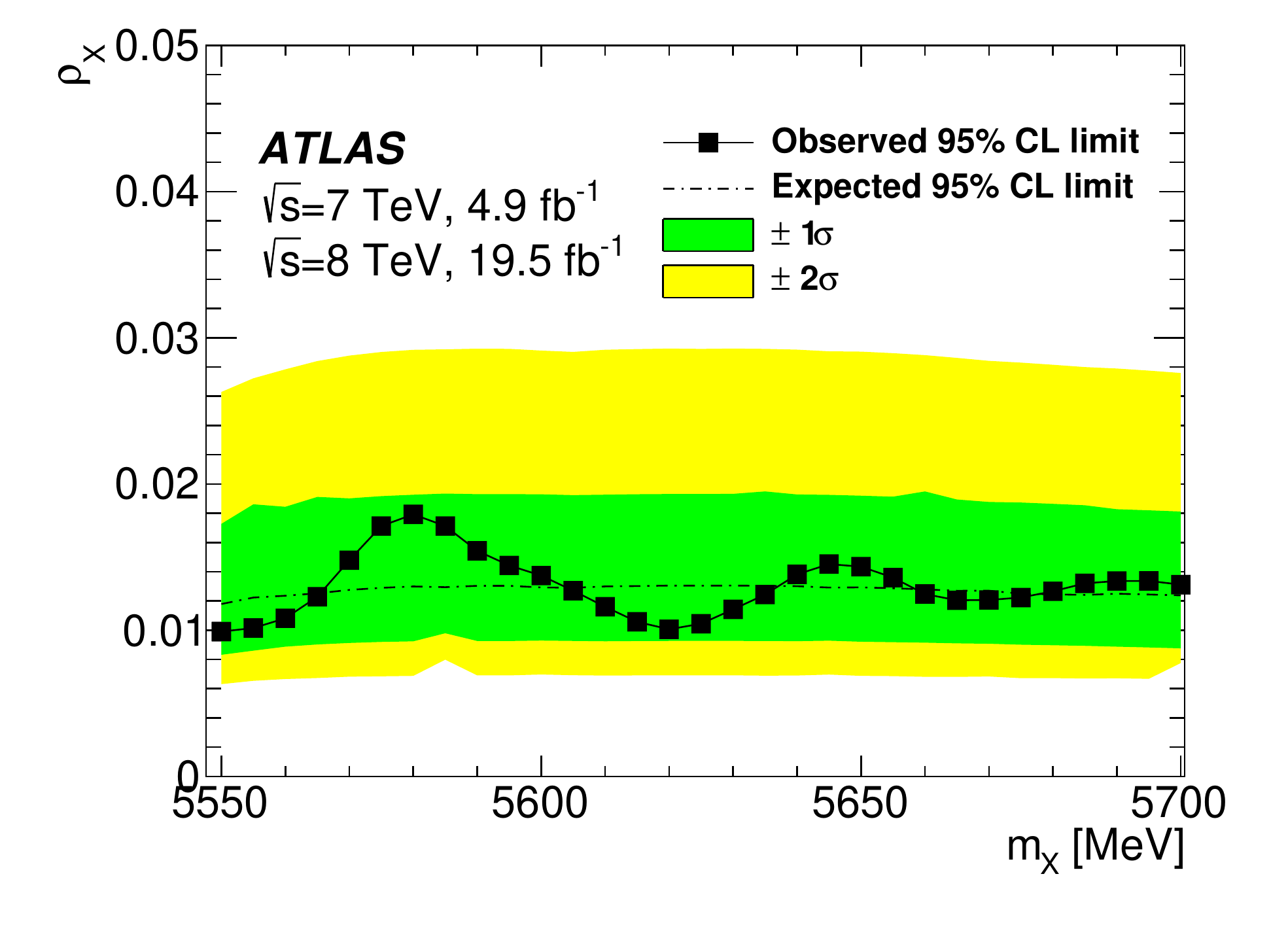}
   
   \caption{ Upper limits on  $\rhoATLAS$  at 95\% CL (black squares connected by line) at different masses of a hypothetical resonant state \textit{X} decaying to \Bspi,  for  events with \cutBspTten.  A BW width of $21.9\pm6.4 (\textrm{stat})^{+5.0}_{-2.5} (\textrm{syst})~\MeV$ is assumed, as reported by D0.  The values include systematic uncertainties.    The expected     95\% CL upper limits (central  black dot-dashed line)  with $\pm 1\sigma$ (green) and $\pm 2 \sigma$ (yellow) uncertainty bands on $\rhoATLAS$ are shown as a function of the assumed resonance mass.}
  \label{Fig:RhoScan}
\end{figure}

As no significant signal is observed, corresponding to the properties of the $X(5568)$ as reported by Ref.~\cite{D0bspi}, upper limits are determined for the number of \Bspi \ signal events, $N(X)$, and for the relative production rate, \rhoATLAS.
These are calculated using the asymptotic approximation
from the profile likelihood formalism~\cite{Cowan:2010js} based on the CL$_\textrm{s}$ frequentist
method~\cite{Read:2002hq}.
To establish the limit  on  the number of \Bspi \ signal events, the PDF models for signal and background, defined respectively by Eqs.~(\ref{Eq:BWSwave}) and~(\ref{Eq:Fbck}),  are used as inputs to the CL$_\textrm{s}$ method. 
Without systematic uncertainties, the extracted  upper limits  at 95\% CL are  $N(X) < 264$  for \cutBspTten \ and  $N(X) < 213$ for  \cutBspTfif. 
Systematic uncertainties affecting these limits are included in the determination of $N(X)$.   
To obtain results that can be compared to the state  $X(5568)$  reported by the D0 Collaboration,  systematic uncertainties are
assigned by  varying  the values of $m_X$ and $\Gamma_X$  independently within Gaussian constraints, with uncertainties equal to those quoted in Ref.~\cite{D0bspi}. 
The  default model of the \textit{X} resonance, which is assumed to be spinless, is changed to a BW $P$-wave resonance.
To  include the systematic uncertainty due to the modeling of the background, the default PDF of Eq.~(\ref{Eq:Fbck})
is replaced by a seventh-order Chebyshev polynomial,
allowing more free parameters in the fit.  For the detector resolution, the default per-candidate mass resolution model is replaced by the sum of three Gaussian functions with a  common mean. The parameters used are determined from the \Bspi \ MC sample. Using these alternative models, upper limits that include systematic uncertainties are extracted, leading to values  \NXten  for \cutBspTten \ and  \NXfif  for  \cutBspTfif. 
To  extract  the upper limits on  $\rhoATLAS$  additional systematic uncertainties  are  included. 
The calculation of $\rhoATLAS$ also depends on the precision of extracting the number of \Bs signal events and  the relative efficiency $\epsilon^{\mathrm{rel}}(X)$. To include these uncertainties, 
the central values and the uncertainties of the number of \Bs signal events and $\epsilon^{\mathrm{rel}}(X)$ are used to construct Gaussian constraints, which are included as additional inputs to the CL$_\textrm{s}$ method. Both the statistical  and systematic uncertainties are included after being summed in quadrature. 
For the \Bs signal, the default fit model of two Gaussian functions is changed to  a triple
Gaussian function and the change in the result is taken as a systematic uncertainty. The uncertainty due to the proper decay time requirement  $t$~> 0.2~ps was estimated by varying  it within the time resolution and found to be negligible.
The resulting upper limits at 95\% CL are  \rhoten for  \cutBspTten \ and \rhofif for   \cutBspTfif.
A hypothesis test is performed for the presence of a \Bspi \ peak for every $5~\MeV$ step in its mass from $5550~\MeV$ to  $5700~\MeV$, assuming a resonant state as described by Eq.~(\ref{Eq:BWSwave}), with a BW width of $21.9~\MeV$~\cite{D0bspi} and \cutBspTten.
For each $\Bspi$ mass tested,  $\epsilon^{\mathrm{rel}}(X)$  is calculated using the same method as for  $X(5568)$. The values of $\epsilon^{\mathrm{rel}}(X)$ vary from 0.50 to 0.55 in the search interval.  The upper limit of $\rhoATLAS$ at 95\% CL is determined for each tested mass. The same systematic uncertainties as in the determination of $\rhoATLAS$ for the state  $X(5568)$  are included, with the exception of the  $X(5568)$ mass uncertainty.
The median expected upper limit at 95\% CL as a function of the \Bspi \ mass is also determined with $\pm 1 \sigma$  and  $\pm 2 \sigma$ error bands. The results are shown in Figure \ref{Fig:RhoScan}.

In conclusion, a search for a new state $X(5568)$ decaying to \Bspi, with properties as reported by the D0
Collaboration, was performed by the ATLAS experiment at the LHC,  using    \ilumiSeven \ of  $pp$ collision data at \CoMEnergySeven \ and    \ilumi \ at \CoMEnergyEight. No significant signal  was found.  Within the acceptance in which this analysis is performed, upper limits on  the number of signal events, $N(X)$,  and  on the $X$ production rate  relative to \Bs mesons, were determined at 95\% CL, resulting in \NXten \ and \rhoten \  for  \cutBspTten, and \NXfif \ and \rhofif \  for   \cutBspTfif.  Limits  are also set for potential \Bspi \ resonances in the mass range from $5550~\MeV$ to  $5700~\MeV$. Across the full range, the upper limit set on $\rhoATLAS$ at
95\% CL varies between 0.010 and 0.018, and does not exceed the $\pm 1\sigma$  error band from the expected limit.




We thank CERN for the very successful operation of the LHC, as well as the
support staff from our institutions without whom ATLAS could not be
operated efficiently.

We acknowledge the support of ANPCyT, Argentina; YerPhI, Armenia; ARC, Australia; BMWFW and FWF, Austria; ANAS, Azerbaijan; SSTC, Belarus; CNPq and FAPESP, Brazil; NSERC, NRC and CFI, Canada; CERN; CONICYT, Chile; CAS, MOST and NSFC, China; COLCIENCIAS, Colombia; MSMT CR, MPO CR and VSC CR, Czech Republic; DNRF and DNSRC, Denmark; IN2P3-CNRS, CEA-DRF/IRFU, France; SRNSFG, Georgia; BMBF, HGF, and MPG, Germany; GSRT, Greece; RGC, Hong Kong SAR, China; ISF, I-CORE and Benoziyo Center, Israel; INFN, Italy; MEXT and JSPS, Japan; CNRST, Morocco; NWO, Netherlands; RCN, Norway; MNiSW and NCN, Poland; FCT, Portugal; MNE/IFA, Romania; MES of Russia and NRC KI, Russian Federation; JINR; MESTD, Serbia; MSSR, Slovakia; ARRS and MIZ\v{S}, Slovenia; DST/NRF, South Africa; MINECO, Spain; SRC and Wallenberg Foundation, Sweden; SERI, SNSF and Cantons of Bern and Geneva, Switzerland; MOST, Taiwan; TAEK, Turkey; STFC, United Kingdom; DOE and NSF, United States of America. In addition, individual groups and members have received support from BCKDF, the Canada Council, CANARIE, CRC, Compute Canada, FQRNT, and the Ontario Innovation Trust, Canada; EPLANET, ERC, ERDF, FP7, Horizon 2020 and Marie Sk{\l}odowska-Curie Actions, European Union; Investissements d'Avenir Labex and Idex, ANR, R{\'e}gion Auvergne and Fondation Partager le Savoir, France; DFG and AvH Foundation, Germany; Herakleitos, Thales and Aristeia programmes co-financed by EU-ESF and the Greek NSRF; BSF, GIF and Minerva, Israel; BRF, Norway; CERCA Programme Generalitat de Catalunya, Generalitat Valenciana, Spain; the Royal Society and Leverhulme Trust, United Kingdom.

The crucial computing support from all WLCG partners is acknowledged gratefully, in particular from CERN, the ATLAS Tier-1 facilities at TRIUMF (Canada), NDGF (Denmark, Norway, Sweden), CC-IN2P3 (France), KIT/GridKA (Germany), INFN-CNAF (Italy), NL-T1 (Netherlands), PIC (Spain), ASGC (Taiwan), RAL (UK) and BNL (USA), the Tier-2 facilities worldwide and large non-WLCG resource providers. Major contributors of computing resources are listed in Ref.~\cite{ATL-GEN-PUB-2016-002}.

\printbibliography


\newpage 
\begin{flushleft}
{\Large The ATLAS Collaboration}

\bigskip

M.~Aaboud$^\textrm{\scriptsize 137d}$,
G.~Aad$^\textrm{\scriptsize 88}$,
B.~Abbott$^\textrm{\scriptsize 115}$,
O.~Abdinov$^\textrm{\scriptsize 12}$$^{,*}$,
B.~Abeloos$^\textrm{\scriptsize 119}$,
S.H.~Abidi$^\textrm{\scriptsize 161}$,
O.S.~AbouZeid$^\textrm{\scriptsize 139}$,
N.L.~Abraham$^\textrm{\scriptsize 151}$,
H.~Abramowicz$^\textrm{\scriptsize 155}$,
H.~Abreu$^\textrm{\scriptsize 154}$
% ...
\bigskip
\\
% ...
$^{12}$ Institute of Physics, Azerbaijan Academy of Sciences, Baku, Azerbaijan\\
% ...
$^{88}$ CPPM, Aix-Marseille Universit{\'e} and CNRS/IN2P3, Marseille, France\\
% ...
$^{115}$ Homer L. Dodge Department of Physics and Astronomy, University of Oklahoma, Norman OK, United States of America\\
% ...
$^{119}$ LAL, Univ. Paris-Sud, CNRS/IN2P3, Universit{\'e} Paris-Saclay, Orsay, France\\
% ...
$^{137}$ $^{(a)}$ Facult{\'e} des Sciences Ain Chock, R{\'e}seau Universitaire de Physique des Hautes Energies - Universit{\'e} Hassan II, Casablanca; $^{(b)}$ Centre National de l'Energie des Sciences Techniques Nucleaires, Rabat; $^{(c)}$ Facult{\'e} des Sciences Semlalia, Universit{\'e} Cadi Ayyad, LPHEA-Marrakech; $^{(d)}$ Facult{\'e} des Sciences, Universit{\'e} Mohamed Premier and LPTPM, Oujda; $^{(e)}$ Facult{\'e} des sciences, Universit{\'e} Mohammed V, Rabat, Morocco\\
% ...
$^{139}$ Santa Cruz Institute for Particle Physics, University of California Santa Cruz, Santa Cruz CA, United States of America\\
% ...
$^{151}$ Department of Physics and Astronomy, University of Sussex, Brighton, United Kingdom\\
% ...
$^{154}$ Department of Physics, Technion: Israel Institute of Technology, Haifa, Israel\\
$^{155}$ Raymond and Beverly Sackler School of Physics and Astronomy, Tel Aviv University, Tel Aviv, Israel\\
% ...
$^{161}$ Department of Physics, University of Toronto, Toronto ON, Canada\\
% ...
$^{a}$ Also at Department of Physics, King's College London, London, United Kingdom\\
$^{b}$ Also at Institute of Physics, Azerbaijan Academy of Sciences, Baku, Azerbaijan\\
$^{c}$ Also at Novosibirsk State University, Novosibirsk, Russia\\
$^{d}$ Also at TRIUMF, Vancouver BC, Canada\\
$^{e}$ Also at Department of Physics {\&} Astronomy, University of Louisville, Louisville, KY, United States of America\\
% ...
$^{*}$ Deceased
\end{flushleft}


\end{document}